\documentclass[12pt,a4paper,english,superscriptaddress,aps]{revtex4}
\usepackage{amssymb}
\usepackage{amsmath} 
\usepackage{slashed}
\usepackage{graphicx}
\usepackage{babel}
\usepackage{hyperref}
\usepackage{color}
\usepackage{comment}
\makeatletter\AtBeginDocument{\let\@elt\relax}\makeatother

\begin{document}

\title{Scattering amplitudes in dimensionless quadratic gravity coupled to QED}

\author{I. F. Cunha}
\email[]{ivana.cunha@icen.ufpa.br} 
\affiliation{Faculdade de F\'{i}sica, Universidade Federal do Par\'{a}, 66075-110, Bel\'{e}m, Par\'a, Brazil}

\author{A. C. Lehum}
\email[]{lehum@ufpa.br}
\affiliation{Faculdade de F\'{i}sica, Universidade Federal do Par\'{a}, 66075-110, Bel\'{e}m, Par\'a, Brazil}

\begin{abstract}
We study ultra-Planckian $2\to2$ scattering in an Abelian gauge theory coupled to agravity, the scale-free and renormalizable realization of quadratic quantum gravity. Focusing on charged fermions and scalars interacting with the photon and the higher-derivative graviton, we present compact analytic expressions for the unpolarized squared matrix elements for a broad set of tree-level processes, including photon--photon, fermion--fermion, fermion--photon, scalar--fermion, scalar--photon, scalar--scalar, and annihilation channels. In contrast to purely graviton-mediated analyses, we retain systematically the photon--graviton interference contributions and verify explicitly the independence of the results on the gravitational gauge-fixing parameter. The amplitudes display characteristic forward/backward enhancements associated with small momentum transfer, amplified by the $1/p^{4}$ graviton propagator, while their high-energy scaling reflects the underlying dimensionless gravitational couplings. Moreover, for all channels analyzed the corresponding differential cross sections exhibit the universal ultra-Planckian scaling $d\sigma/d\Omega \propto 1/s$, where $s$ is the Mandelstam invariant (the squared center-of-momentum energy). Our results furnish a unified amplitude-level description of how higher-derivative gravity reshapes familiar QED scattering at ultra-Planckian energies and provide analytic building blocks for further studies of IR definitions and UV consistency in agravity with matter.
\end{abstract}

\maketitle

\section{Introduction}

Tree-level scattering amplitudes provide a particularly sharp probe of UV completions of gravity: they make explicit the momentum dependence imposed by the underlying degrees of freedom and, in higher-derivative theories, they expose the cancellations that must occur if unitarity is to be realized in a nonstandard way. In this paper we present a systematic set of $2\to2$ scattering amplitudes in agravity coupled to QED with scalar and fermionic matter, including not only the pure graviton-mediated contributions but also the interference terms between photon-mediated and graviton-mediated exchanges.  Our aim is twofold: to supply a compact set of analytic results that can be used as building blocks in phenomenological and conceptual studies, and to assess, in explicit channels involving matter, how the characteristic higher-derivative dynamics manifests itself in angular dependence, IR enhancements, and UV scaling.

The broader setting is familiar.  Quantized around flat spacetime, Einstein's General Relativity leads to a nonrenormalizable perturbative expansion \cite{tHooft:1974toh,Deser:1974zzd,Deser:1974cy}, which may be interpreted as an effective field theory below the Planck scale \cite{Donoghue:1994dn,Burgess:2003jk,Donoghue:2017pgk}.  A distinct approach is provided by quadratic quantum gravity, in which curvature-squared terms render the theory power-counting renormalizable \cite{Stelle:1976gc,Tomboulis:1977jk,Odintsov:1991nd,Salvio:2014soa,Salvio:2017qkx,Einhorn:2014gfa,Buchbinder:2017lnd,Salvio:2018crh,Donoghue:2021cza}. Agravity realizes the scale-free version of this idea by retaining only quadratic curvature invariants in the classical action \cite{Salvio:2014soa,Aoki:2021skm,Aoki:2024jhr,Alvarez-Luna:2022hka}, so that the graviton propagator behaves as $1/p^{4}$ at large momentum \cite{Buoninfante:2023ryt}, with the possibility that an effective Planck scale emerges dynamically at the quantum level \cite{Gialamas:2020snr,Salvio:2020axm}.  The unitarity question in higher-derivative gravity has been revisited in recent years \cite{Anselmi:2017ygm,Donoghue:2019fcb,Salvio:2019wcp,Salvio:2024joi}, and explicit amplitude studies have revealed distinctive IR enhancements already at tree level in channels involving matter and gauge fields \cite{Salvio:2018kwh}.  From this perspective, it is valuable to examine concrete scattering processes in which higher-derivative gravity competes and interferes with standard gauge interactions.

The present work develops a parallel amplitude analysis in the Abelian sector, in the same spirit as Ref.~\cite{Cunha:2025djg}, and is motivated by the amplitude-based program initiated by Holdom \cite{Holdom:2021oii,Holdom:2021hlo}. We focus on QED coupled to agravity with charged scalars and fermions and compute the relevant tree-level $2\to2$ channels, including the photon--graviton interference terms that are absent in a purely graviton-mediated treatment. Holdom showed that, in quadratic gravity, purely photonic processes can exhibit well-behaved differential cross sections and nontrivial UV scaling, highlighting the subtleties in the relation between unitarity and positivity in a higher-derivative setting. The channels considered here extend that perspective to reactions involving matter fields, providing a clean arena---by virtue of the universal gravitational coupling to the energy--momentum tensor---to assess how higher-derivative dynamics reshapes familiar QED amplitudes, both in the collinear regime of small momentum transfer and in the deep UV.

The remainder of the paper is organized as follows.  In Sec.~\ref{sec02} we present the model and collect the propagators and interaction vertices needed for the tree-level computations. Section~\ref{sec03} contains the evaluation of the graviton-mediated and photon-mediated amplitudes for the relevant $2\to2$ processes, including their interference, together with a discussion of the resulting angular structures and UV behavior.  Our conclusions are summarized in Sec.~\ref{summary}.  Throughout we work in natural units $c=\hbar=1$.

\section{The QED-agravity Lagrangian}\label{sec02}

We consider an Abelian gauge theory with charged scalar and fermion matter coupled to four-derivative, dimensionless gravity (``agravity''). The classical action is taken to be
\begin{eqnarray}
\label{eq01}
\mathcal{S}
&=&
\int d^{4}x\,\sqrt{-g}\,
\Bigg\{
\frac{R^{2}}{6f_{0}^{2}}
+\frac{1}{f_{2}^{2}}
\left(\frac{1}{3}R^{2}-R_{\mu\nu}R^{\mu\nu}\right)
-\frac{1}{4}F_{\mu\nu}F^{\mu\nu}
\nonumber\\
&&\hspace{1.15cm}
+i\,\bar\psi\,\gamma^{\mu}\big(\nabla_{\mu}-ieA_{\mu}\big)\psi
+\big(D_{\mu}\phi\big)^{\dagger}D^{\mu}\phi
-\frac{\lambda}{4}\,(\phi^{\dagger}\phi)^{2}
+\xi\,\phi^{\dagger}\phi\,R
\Bigg\},
\end{eqnarray}
where $R$ and $R_{\mu\nu}$ are the Ricci scalar and Ricci tensor, $\nabla_{\mu}$ is the gravitational covariant derivative acting on spinors, and
\begin{equation}
F_{\mu\nu}=\partial_{\mu}A_{\nu}-\partial_{\nu}A_{\mu},
\qquad
D_{\mu}\phi=(\partial_{\mu}-ieA_{\mu})\phi
\end{equation}
denote, respectively, the Abelian field strength and the gauge-covariant derivative of the scalar. The parameters $f_{0}$ and $f_{2}$ are the dimensionless gravitational couplings controlling the $R^{2}$ and Weyl-squared sectors (here written in the equivalent Ricci-basis $\frac{1}{3}R^{2}-R_{\mu\nu}R^{\mu\nu}$), while $e$ and $\lambda$ are the usual Abelian gauge and quartic-scalar couplings. Finally, $\xi$ is the nonminimal scalar-curvature coupling.  Because the action contains no Einstein--Hilbert term and no explicit mass parameters, it is classically scale-free; in particular, the interaction $\xi\,\phi^{\dagger}\phi\,R$ respects this scaleless
structure, since $\xi$ is dimensionless in four spacetime dimensions.

We work in a tetrad formulation for the fermionic sector. The curved-space Dirac matrices are defined by
\begin{equation}
\gamma^{\mu}(x)=e^{\mu}{}_{\alpha}(x)\,\gamma^{\alpha},
\qquad
g_{\mu\nu}(x)=e_{\mu}{}^{\alpha}(x)\,e_{\nu}{}^{\beta}(x)\,\eta_{\alpha\beta},
\end{equation}
with $\eta_{\alpha\beta}=\mathrm{diag}(+,-,-,-)$. Spinor covariant derivatives are written as
\begin{equation}
\overrightarrow{\nabla}_{\mu}\psi=\big(\partial_{\mu}+i\,\omega_{\mu}\big)\psi,
\qquad
\bar\psi\,\overleftarrow{\nabla}_{\mu}=\partial_{\mu}\bar\psi-i\,\bar\psi\,\omega_{\mu},
\end{equation}
where the spin connection is $\omega_{\mu}\equiv \frac{1}{4}\,\omega_{\mu}{}^{\alpha\beta}\, \sigma_{\alpha\beta}$, with $\sigma^{\alpha\beta}=\frac{i}{2}[\gamma^{\alpha},\gamma^{\beta}]$.  For completeness, $\omega_{\mu}{}^{\alpha\beta}$ may be expressed in terms of the vierbein as
\begin{eqnarray}
\omega_{\mu}{}^{\alpha\beta}
&=&
\frac{1}{2}\,e^{\nu\alpha}\big(\partial_{\mu}e_{\nu}{}^{\beta}-\partial_{\nu}e_{\mu}{}^{\beta}\big)
-\frac{1}{2}\,e^{\nu\beta}\big(\partial_{\mu}e_{\nu}{}^{\alpha}-\partial_{\nu}e_{\mu}{}^{\alpha}\big)
\nonumber\\
&&
-\frac{1}{2}\,e^{\rho\alpha}e^{\sigma\beta}
\big(\partial_{\rho}e_{\sigma\gamma}-\partial_{\sigma}e_{\rho\gamma}\big)e_{\mu}{}^{\gamma},
\label{eq:spin_connection}
\end{eqnarray}
which is equivalent to the standard torsionless Levi--Civita expression.  In what follows, indices $\mu,\nu,\ldots$ refer to spacetime coordinates, while $\alpha,\beta,\ldots$ label tangent-space components.

To compute the tree-level $2\to2$ amplitudes we expand about flat spacetime,
\begin{equation}
g_{\mu\nu}=\eta_{\mu\nu}+h_{\mu\nu}\qquad\text{(exact)},
\qquad
g^{\mu\nu}=\eta^{\mu\nu}-h^{\mu\nu}+\mathcal{O}(h^{2}),
\label{eq:metric_expansion}
\end{equation}
and retain only the interactions linear in $h_{\mu\nu}$, corresponding to the one-graviton exchange approximation.  In this approximation Faddeev--Popov ghosts do not contribute, but gauge-fixing is still required to define propagators.  For the graviton we employ the de Donder-type gauge-fixing term
\begin{equation}
\mathcal{L}_{\mathrm{GF},h}
=
-\frac{1}{2\zeta_{g}}
\Big[\partial^{\nu}\Big(h_{\mu\nu}-\frac{1}{2}\eta_{\mu\nu}h\Big)\Big]
\Big[\partial^{\alpha}\Big(h^{\mu}{}_{\alpha}-\frac{1}{2}\eta^{\mu}{}_{\alpha}h\Big)\Big],
\label{eq:gf_graviton}
\end{equation}
and for the Abelian gauge field we choose the covariant gauge
\begin{equation}
\mathcal{L}_{\mathrm{GF},A}
=
-\frac{1}{2\xi}\,(\partial_{\mu}A^{\mu})^{2},
\label{eq:gf_photon}
\end{equation}
with gauge parameters $\zeta_{g}$ and $\xi$.  Explicit linearized Lagrangians for matter coupled to quadratic gravity, including non-Abelian generalizations, may be found in
Refs.~\cite{Choi:1994ax,Souza:2022ovu,Souza:2023wzv,Gomes:2024onm}.

From the quadratic action one obtains the gauge-field propagator
\begin{equation}
\Delta_{\mu\nu}(p)
=
-\frac{i}{p^{2}}
\left(T_{\mu\nu}+\xi\,L_{\mu\nu}\right),
\label{eq:prop_gauge}
\end{equation}
and the graviton propagator in agravity,
\begin{equation}
\Delta_{\mu\nu\rho\sigma}(p)
=
\frac{i}{p^{4}}
\left[
-2f_{2}^{2}\,P^{(2)}_{\mu\nu\rho\sigma}
+f_{0}^{2}\,P^{(0)}_{\mu\nu\rho\sigma}
+2\zeta_{g}\!\left(P^{(1)}_{\mu\nu\rho\sigma}+\frac{1}{2}P^{(0w)}_{\mu\nu\rho\sigma}\right)
\right],
\label{eq:prop_graviton}
\end{equation}
\noindent The tensor structures are built from the transverse and longitudinal projectors,
\begin{equation}
T_{\mu\nu}=\eta_{\mu\nu}-\frac{p_{\mu}p_{\nu}}{p^{2}},
\qquad
L_{\mu\nu}=\frac{p_{\mu}p_{\nu}}{p^{2}},
\label{eq:TL_projectors}
\end{equation}
and the standard spin projectors in the space of symmetric rank-2 tensors,
\begin{eqnarray}
P^{(2)}_{\mu\nu\rho\sigma}
&=&
\frac{1}{2}\Big(T_{\mu\rho}T_{\nu\sigma}+T_{\mu\sigma}T_{\nu\rho}\Big)
-\frac{1}{D-1}\,T_{\mu\nu}T_{\rho\sigma},
\nonumber\\
P^{(1)}_{\mu\nu\rho\sigma}
&=&
\frac{1}{2}\Big(
T_{\mu\rho}L_{\nu\sigma}+T_{\mu\sigma}L_{\nu\rho}
+L_{\mu\rho}T_{\nu\sigma}+L_{\mu\sigma}T_{\nu\rho}
\Big),
\nonumber\\
P^{(0)}_{\mu\nu\rho\sigma}
&=&
\frac{1}{D-1}\,T_{\mu\nu}T_{\rho\sigma},
\qquad
P^{(0w)}_{\mu\nu\rho\sigma}=L_{\mu\nu}L_{\rho\sigma},
\label{eq:spin_projectors}
\end{eqnarray}
which satisfy the usual completeness relations in the symmetric tensor space.

With these conventions fixed, we proceed in the next section to the explicit evaluation of the tree-level scattering amplitudes in agravity--QED, including both the graviton-mediated contributions and their interference with photon exchange.

\section{Agravity-Mediated Scattering Processes}
\label{sec03}

In this section we present the tree-level $2\to2$ scattering amplitudes induced by the spin-2 and spin-0 sectors of agravity, including their interference with the corresponding QED contributions whenever photon exchange is present.  Since our primary goal is to provide a transparent catalog of analytic results and to expose their angular and energy dependence, it is indeed convenient to discuss the channels \emph{process by process}.

Before proceeding, it is useful to relate the present Abelian results to our previous ultra-Planckian study in the non-Abelian sector, Ref.~\cite{Cunha:2025djg}. The setup and the linearized agravity propagator are the same in both works; consequently, whenever a given QED process is mediated purely by graviton exchange, its kinematic dependence follows the same higher-derivative pattern found in Ref.~\cite{Cunha:2025djg}, up to the replacement of color structures by Abelian charge factors and the absence of non-Abelian self-interactions. What is new here is the systematic inclusion of photon exchange and, crucially, the photon--graviton interference terms, which are absent in Ref.~\cite{Cunha:2025djg} by construction. For the reader's convenience, we note the following correspondence between representative results: the purely gravitational gauge-boson channel in Ref.~\cite{Cunha:2025djg} [Eq.~(12)] is echoed here by the purely gravitational $\gamma\gamma\to\gamma\gamma$ result in Eq.~\eqref{eq:M2_photon_avg}, while the identical-fermion channel in Ref.~\cite{Cunha:2025djg} [Eq.~(25)] finds its Abelian counterpart in Eq.~\eqref{scat_eeee}, now augmented by the QED contribution and its interference with graviton exchange. In addition, the mixed matter--gauge channels discussed in Ref.~\cite{Cunha:2025djg} (e.g. gq scattering) are structurally analogous to the crossed-channel relations implemented below for Compton scattering, Eq.~\eqref{eq:M2_eg_to_eg}, though the present amplitudes also contain photon-mediated pieces.

The characteristic regime in which the higher-derivative gravitational dynamics becomes prominent is the ultra-Planckian domain, $E\gg M_{P}$, where the four-derivative graviton propagator governs the momentum dependence of the exchange diagrams. Throughout, we keep the gravitational gauge-fixing parameter $\zeta_{g}$ arbitrary and verify explicitly that the final squared amplitudes are independent of $\zeta_{g}$, as required by diffeomorphism gauge invariance at tree level.

We start with electron--electron scattering, $e^{-}e^{-}\to e^{-}e^{-}$, which provides the simplest setting in which to exhibit the interplay between photon exchange and the gravitational contribution. The calculations were performed with standard \texttt{Mathematica} packages~\cite{feyncalc,Hahn:1998yk,feynarts,feynrules,feynrules1,feyncalc1,feynhelpers,feyncalc2}.

\subsection{Electron-electron scattering into two electrons}

\begin{figure}[h!]
	\includegraphics[angle=0 ,width=13.5cm]{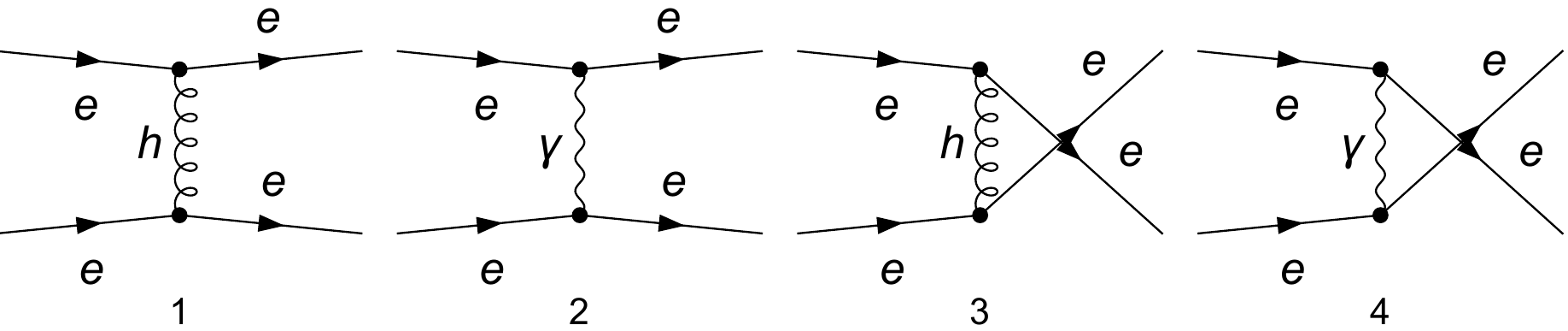}
    \caption{\label{fig:ee_scattering}
Tree-level diagrams contributing to $e^-e^- \to e^-e^-$.
Straight, wavy, and curly lines denote, respectively, the electron, photon, and graviton propagators.}
	\label{fig01}
\end{figure}

The squared matrix element for elastic electron--electron scattering,
$e^{-}e^{-}\to e^{-}e^{-}$, mediated by photon exchange and by the agravity interaction (see Fig.~\ref{fig01}), can be written,
in terms of the Mandelstam variables $t$ and $u$, as
\begin{eqnarray}
\label{scat_eeee}
\overline{|\mathcal{M}|^{2}}
&=&
\frac{4\,(t^{2}+t u+u^{2})^{2}e^{4}}{t^{2}u^{2}}
-\frac{f_{2}^{2}\Big(4 t^{6}+8 t^{5}u+8 t^{4}u^{2}+7 t^{3}u^{3}+8 t^{2}u^{4}+8 t u^{5}+4 u^{6}\Big)e^{2}}{2\,t^{3}u^{3}}
\\[2pt]
&&
+\frac{f_{2}^{4}\Big(16 t^{8}+32 t^{7}u+25 t^{6}u^{2}+30 t^{5}u^{3}+43 t^{4}u^{4}
+30 t^{3}u^{5}+25 t^{2}u^{6}+32 t u^{7}+16 u^{8}\Big)}{64\,t^{4}u^{4}} \, .\nonumber
\end{eqnarray}
The above equation also provides a useful point of contact with the identical-fermion channel discussed in Ref.~\cite{Cunha:2025djg}. In particular, the purely gravitational contribution, proportional to $f_{2}^{4}$, displays the same higher-derivative \emph{kinematic pattern}---namely, the characteristic $t$- and $u$-channel pole structure and the associated forward/backward enhancement---as in the graviton-mediated analysis of Ref.~\cite{Cunha:2025djg}. The overall numerical coefficients are not expected to coincide term by term, since they depend on the particle content and, in the non-Abelian case, on the corresponding color structures. The present Abelian result contains, in addition, the standard QED term $\propto e^{4}$ and the photon--graviton interference $\propto e^{2}f_{2}^{2}$, which are absent in Ref.~\cite{Cunha:2025djg}.

Here $\overline{|\mathcal{M}|^{2}}$ denotes the unpolarized squared amplitude, i.e.,
the matrix element squared summed over final-state spins and averaged over the
initial-state spins,
\begin{equation}
\overline{|\mathcal{M}|^{2}}
=
\frac{1}{4}\,|\mathcal{M}|^{2}_{\rm spin\text{-}summed}\, .
\end{equation}

In order to investigate the conditions under which the squared amplitude $|\mathcal{M}|^2$ is positive, we evaluate it in the center-of-momentum (CoM) frame. In this frame, the Mandelstam variables are given by
\begin{equation}
t=-\frac{s}{2}(1-\cos\theta)=-s\,\sin^{2}\!\frac{\theta}{2},
\qquad
u=-\frac{s}{2}(1+\cos\theta)=-s\,\cos^{2}\!\frac{\theta}{2}.
\label{eq:stu_com_massless}
\end{equation}
Substituting these expressions into the amplitude, we obtain
\begin{eqnarray}\label{m2eeeetheta}
\overline{|\mathcal{M}|^{2}}_{\rm CoM}(\theta)
&=&
\frac{\csc^{8}\!\theta}{64}\Bigg[64\,e^{4}\,(7+\cos 2\theta)^{2}\sin^{4}\!\theta\nonumber\\
&&-\;f_{2}^{2}\,e^{2}\Big(4662+3343\cos 2\theta+186\cos 4\theta+\cos 6\theta\Big)\sin^{2}\!\theta\nonumber\\
&&+
f_{2}^{4}\Big(249+1700\cos^{2}\!\theta+1998\cos^{4}\!\theta+148\cos^{6}\!\theta+\cos^{8}\!\theta\Big)
\Bigg] .
\end{eqnarray}

The expression \eqref{m2eeeetheta} is strictly positive for all scattering angles $0<\theta<\pi$ and for arbitrary real couplings $e$ and $f_2$. 

The collinear structure is entirely contained in the prefactor of the amplitude, proportional to $\csc^{8}\!\theta$, leading to the behaviour $\overline{|{\mathcal{M}}|^2} \sim 1/\sin^8\theta$ near $\theta \to 0,\pi$. This behaviour is the expected enhancement from spin-2 exchange in the $t$- and $u$-channels~\cite{Salvio:2018kwh}. In the limit where $f_2$ becomes asymptotically small (as expected from asymptotic freedom~\cite{Salvio:2014soa}), this enhancement is suppressed and the angular distribution is governed solely by the electromagnetic contribution.

\subsection{Photon-photon scattering into two photons}

\begin{figure}[h!]
	\includegraphics[angle=0 ,width=11.5cm]{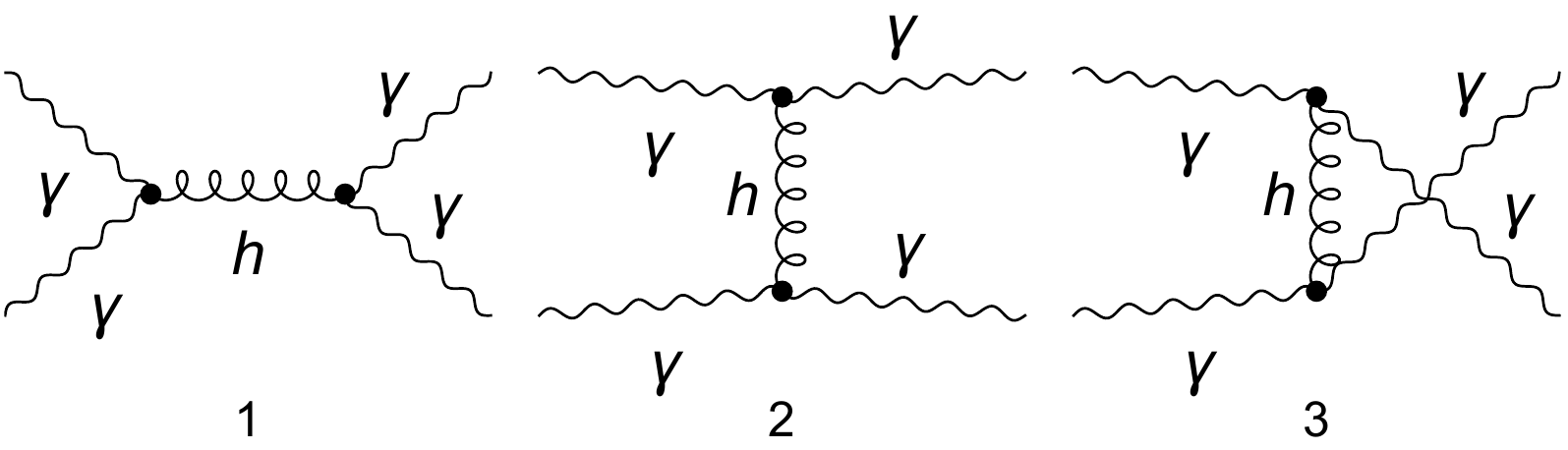}
	\caption{Tree-level graviton-exchange diagrams for $\gamma\gamma \to \gamma\gamma$.}
	\label{fig02}
\end{figure}

We now consider the process $\gamma \gamma \rightarrow \gamma \gamma$ mediated by a graviton exchange shown in Fig.~\ref{fig02}. After summing the squared amplitude over the physical photon polarizations (helicities) in the
final state and averaging over the initial-state photon polarizations, i.e.
\begin{equation}
\overline{|\mathcal{M}|^{2}}
=
\frac{1}{4}\sum_{\lambda_1,\lambda_2}\sum_{\lambda_3,\lambda_4}
\left|\mathcal{M}_{\lambda_1\lambda_2\to\lambda_3\lambda_4}\right|^{2},
\label{eq:polavg_photon}
\end{equation}
we obtain the following unpolarized squared matrix element (expressed in terms of the Mandelstam
invariants $s,t,u$):
\begin{equation}
\overline{|\mathcal{M}|^{2}}
=
\frac{f_{2}^{4}}{8\,s^{4}t^{4}u^{4}}
\Big[
s^{8}\,(t^{2}+u^{2})^{2}
+t^{4}u^{4}\,(t^{4}+u^{4})
+2s^{2}t^{2}u^{2}\,(t^{6}+u^{6})
+s^{4}\,(t^{8}+u^{8})
\Big] \, .
\label{eq:M2_photon_avg}
\end{equation}
Here $f_{2}$ is the dimensionless agravity coupling associated with the spin-2 sector, and the bar denotes the standard unpolarized average over the $2\times2$ physical polarization states of the massless photons in the initial state. This result may be compared with the graviton-mediated gauge-boson channel discussed in Ref.~\cite{Cunha:2025djg}, as it exhibits the characteristic higher-derivative forward/backward enhancement of quadratic gravity. For on-shell photons the spin-0 sector does not contribute, so the pure-gravity result depends only on the spin-2 coupling $f_{2}$.

In the CoM frame, the unpolarized squared matrix element depends only on the scattering angle $\theta$. After summing over the final-state photon helicities and averaging over the initial-state helicities, the result can be written as
\begin{eqnarray}
\overline{|\mathcal{M}|^{2}}_{\rm CoM}(\theta)
&=&
\frac{f_{2}^{4}}{16}\,
\csc^{8}\!\theta\,
\Big(
537
+1578\cos^{2}\!\theta
+1503\cos^{4}\!\theta
+364\cos^{6}\!\theta
\nonumber\\
&&\qquad \qquad +103\cos^{8}\!\theta
+10\cos^{10}\!\theta
+\cos^{12}\!\theta
\Big)\,,
\label{eq:M2_CoM_photons}
\end{eqnarray}
where the bar denotes the standard unpolarized average over the $2\times 2$ physical polarization states of the incoming massless photons. Equation~\eqref{eq:M2_CoM_photons} exhibits a strong enhancement in the forward and backward directions~\cite{Holdom:2021oii} through the factor $\csc^{8}\!\theta$. This behavior is the angular manifestation of the singular kinematics associated with small momentum transfer in the crossed-channel regions: using the massless CoM relations of Eq.~\eqref{eq:stu_com_massless}, one finds that $\theta\to0$ corresponds to $t\to0$ (with
$u\to -s$), whereas $\theta\to\pi$ corresponds to $u\to0$ (with $t\to -s$).

\subsection{Electron-positron annihilation into two photons and Compton scattering}

\begin{figure}[h!]
	\includegraphics[angle=0 ,width=11.5cm]{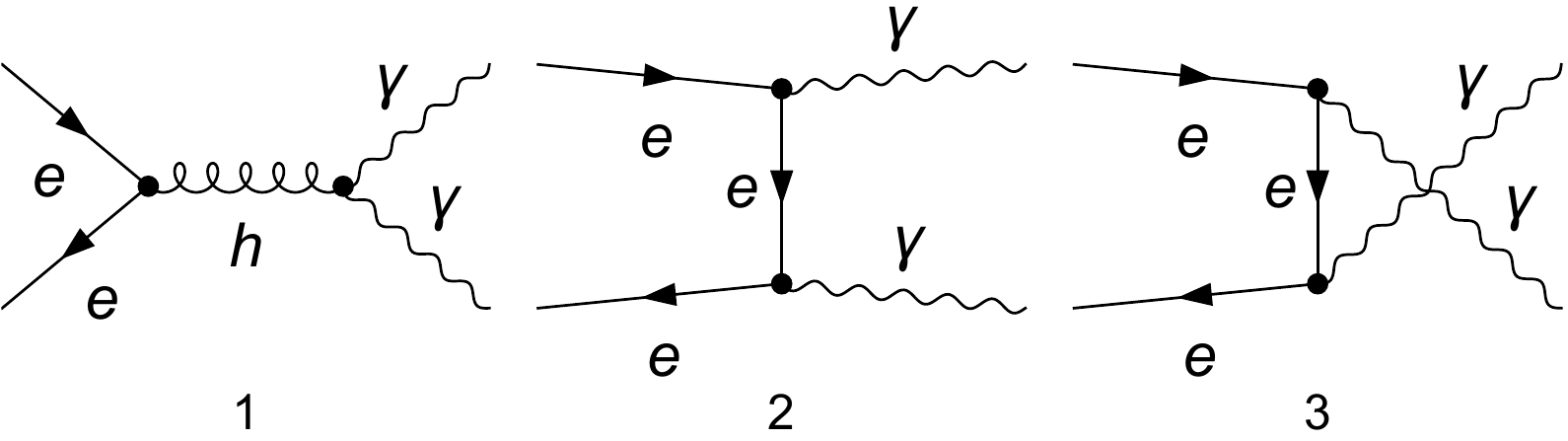}
	\caption{Tree-level diagrams for $e^-e^+ \to \gamma\gamma$, including the standard QED contributions and the additional graviton-exchange diagram.}
	\label{fig03}
\end{figure}

For the process $e^{-}e^{+}\to\gamma\gamma$, Fig. \ref{fig03}, the unpolarized squared matrix element is defined by summing over the physical photon helicities in the final state and averaging over the spins of the
incoming electron and positron,
\begin{equation}
\overline{|\mathcal{M}|^{2}}
=
\frac{1}{4}\sum_{s_{1},s_{2}}
\sum_{\lambda_{1},\lambda_{2}}
\left|\mathcal{M}_{\,s_{1}s_{2}\to \lambda_{1}\lambda_{2}}\right|^{2}.
\label{eq:polavg_epgg}
\end{equation}
In terms of the Mandelstam invariants $s,t,u$, we obtain
\begin{equation}
\overline{|\mathcal{M}|^{2}}
=
\frac{(t^{2}+u^{2})\big(f_{2}^{2}\,t\,u+4\,e^{2}\,s^{2}\big)^{2}}
{8\,s^{4}\,t\,u}\, .
\label{eq:M2_ep_to_gg}
\end{equation}
In contrast with Ref.~\cite{Cunha:2025djg}, where gauge-boson internal lines were not included, the compact squared structure in Eq.~\eqref{eq:M2_ep_to_gg} makes the photon--graviton interference explicit at the amplitude level, providing a clean diagnostic of how the spin-2 sector reshapes a standard QED annihilation channel at ultra-Planckian energies.

Several features of Eq.~\eqref{eq:M2_ep_to_gg} are immediate. First, the factor $(t^{2}+u^{2})$ encodes the familiar helicity/kinematic structure of $e^{+}e^{-}\to\gamma\gamma$ and makes the result symmetric under $t\leftrightarrow u$, as required by Bose symmetry of the two-photon final state. Second, the combination $(f_{2}^{2}\,t\,u+4s^{2}e^{2})$ reveals the interference between the electromagnetic amplitude and the spin-2 agravity exchange: the $4s^{2}e^{2}$ term reproduces the pure QED contribution, the $f_{2}^{2}tu$ term represents the leading agravity correction, and the square exhibits how the two contributions combine at the amplitude level. Third, the overall factor $1/(t\,u)$ signals the expected enhancement in the collinear limits $t\to0$ or $u\to0$, corresponding to forward/backward emission of one of the photons. These singular limits are the standard remnant of massless kinematics in $2\to2$ scattering and, in practice, are regulated by experimental angular cuts or by keeping a small fermion mass when forming inclusive observables.

In the CoM frame, the unpolarized squared matrix element for $e^{+}e^{-}\to\gamma\gamma$ depends only on the scattering angle $\theta$ between the incoming electron and one of the outgoing photons. After summing over the physical photon helicities in the final state and averaging over the spins of the initial $e^{+}e^{-}$ pair, one finds
\begin{equation}
\overline{|\mathcal{M}|^{2}}_{\rm CoM}(\theta)
=
\frac{1+\cos^{2}\!\theta}{64}
\left(
f_{2}^{2}\,\sin\theta
+16\,e^{2}\,\csc\theta
\right)^{2}.
\label{eq:M2_epgg_CoM}
\end{equation}
The prefactor $1+\cos^{2}\!\theta$ is the characteristic angular dependence of $e^{+}e^{-}\to\gamma\gamma$ in the ultrarelativistic regime, reflecting the helicity structure of the annihilation into two transverse photons and guaranteeing the expected symmetry under $\theta\to\pi-\theta$ (equivalently, $t\leftrightarrow u$). The quantity in parentheses displays the superposition of the electromagnetic contribution and the spin-2 agravity correction: the term proportional to $e^{2}$ reproduces the QED amplitude, while the term proportional to $f_{2}^{2}$ encodes the leading effect of the agravity sector. The appearance of $\csc\theta$ makes explicit the enhancement in the forward/backward directions, $\theta\to0$ and $\theta\to\pi$, which corresponds to the collinear kinematics where one of the Mandelstam invariants $t$ or $u$ approaches zero. As in any massless $2\to2$ process mediated by long-range interactions, integrated observables must be defined with an angular cut (or, equivalently, with a cut on $|t|$ and $|u|$) set by experimental
resolution or by the appropriate IR physics.

\begin{figure}[h!]
	\includegraphics[angle=0 ,width=11.5cm]{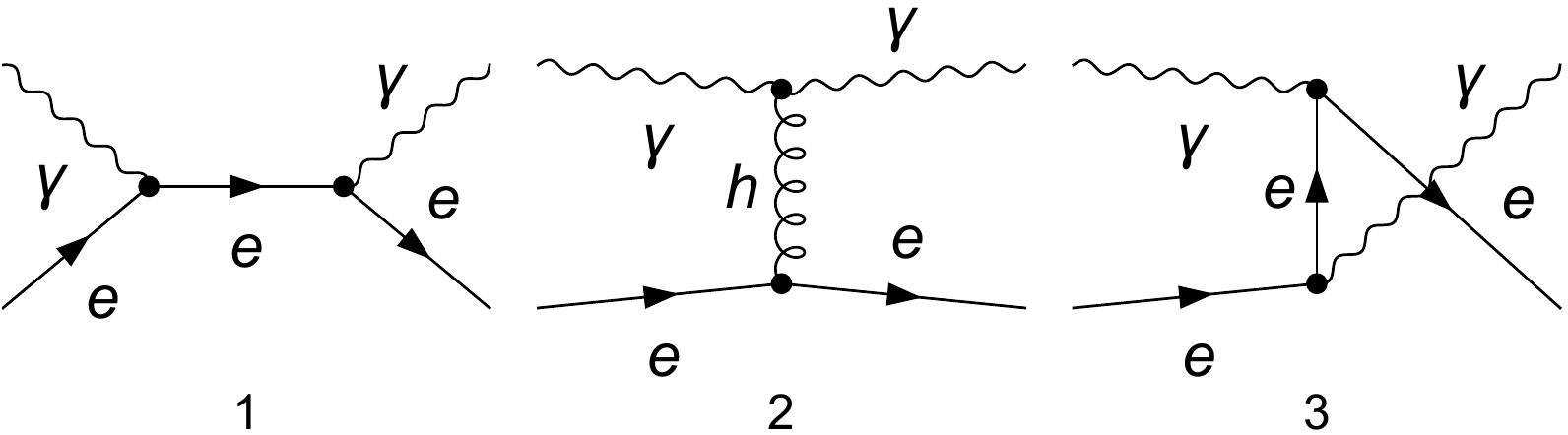}
	\caption{Tree-level diagrams for Compton scattering $e^-\gamma \to e^-\gamma$, obtained from Fig.~\ref{fig03} by crossing.}
	\label{fig04}
\end{figure}

The Compton process $e^{-}\gamma\to e^{-}\gamma$, Fig. \ref{fig04}, is obtained from the annihilation channel $e^{-}e^{+}\to\gamma\gamma$ by crossing symmetry. Using the Mandelstam invariants
\begin{equation}
\label{eq:stu_annih}
s=(p_{1}+p_{2})^{2},\qquad
t=(p_{1}-p_{3})^{2},\qquad
u=(p_{1}-p_{4})^{2},
\end{equation}
for $e^{-}(p_{1})\,e^{+}(p_{2})\to\gamma(p_{3})\,\gamma(p_{4})$, and
\begin{equation}
\label{eq:stu_compton}
s_{C}=(p_{1}+p_{3})^{2},\qquad
t_{C}=(p_{1}-p_{3})^{2}=(p_{3}-p_{4})^{2},\qquad
u_{C}=(p_{1}-p_{4})^{2},
\end{equation}
for $e^{-}(p_{1})\,\gamma(p_{3})\to e^{-}(p_{3})\,\gamma(p_{4})$, the two channels are related by the
crossing permutation
\begin{equation}
\label{eq:crossing_st}
(s,t,u)\;\longrightarrow\;(t_{C},s_{C},u_{C})\equiv (t,s,u)\,,
\end{equation}
i.e.\ by exchanging $s\leftrightarrow t$ while keeping $u$ fixed (within our momentum-labeling convention). Consequently, the unpolarized squared amplitude for Compton scattering can be obtained from the corresponding result for $e^{-}e^{+}\to\gamma\gamma$ by the crossing replacement $s\leftrightarrow t$. This yields
\begin{equation}
\overline{|\mathcal{M}|^{2}}
=
\frac{(s^{2}+u^{2})\big(f_{2}^{2}\,s\,u+4\,e^{2}\,t^{2}\big)^{2}}
{8\,t^{4}\,s\,u}\,,
\label{eq:M2_eg_to_eg}
\end{equation}
where the bar denotes the standard sum over final-state photon helicities together with the average over the initial-state electron spin and photon helicity. The crossing permutation in Eq.~\eqref{eq:M2_eg_to_eg} plays here a role analogous to the kinematic substitutions used in Ref.~\cite{Cunha:2025djg} to relate different channels within a given process, but now it acts on amplitudes that include both graviton- and photon-mediated contributions.

In the CoM frame, where the kinematics is conveniently parametrized by the scattering angle $\theta$, Eq.~\eqref{eq:M2_eg_to_eg} reduces to
\begin{equation}
\overline{|\mathcal{M}|^{2}}_{\rm CoM}(\theta)
=
\frac{\big(5+\cos\theta\,(2+\cos\theta)\big)}
{4\,(\cos\theta-1)^{4}\,(1+\cos\theta)}\,
\Big[
f_{2}^{2}\,(1+\cos\theta)
-8\,e^{2}\,\sin^{4}\!\Big(\frac{\theta}{2}\Big)
\Big]^{2}.
\label{eq:M2_eg_to_eg_CoM}
\end{equation}
The angular dependence in Eq.~\eqref{eq:M2_eg_to_eg_CoM} makes explicit the expected enhancement in the forward direction, $\theta\to 0$, which corresponds to vanishing momentum transfer $t\to 0$. In particular, the factor $(\cos\theta-1)^{-4}$ reflects the strong small-angle behavior associated with $t$-channel exchange in the massless limit, while the remaining $(1+\cos\theta)^{-1}$ structure encodes the complementary collinear configuration of the crossed channel. In practice, integrated observables are defined with the usual experimental angular cuts (equivalently, cuts on $|t|$), which regulate these kinematic singularities.

\subsection{Charged scalar-electron Scattering}

\begin{figure}[h!]
	\includegraphics[angle=0 ,width=7cm]{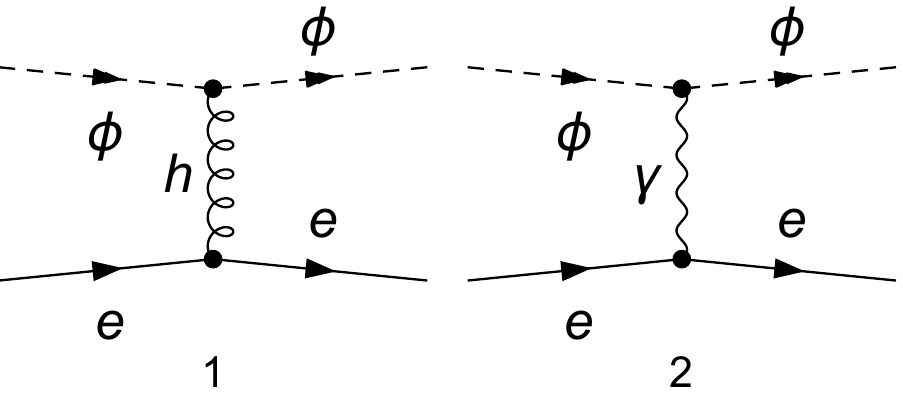}
\caption{Tree-level diagrams contributing to $\phi\,e^- \to \phi\,e^-$ (charged scalar--electron scattering), including photon exchange and graviton exchange.
Straight, wavy, and curly lines denote, respectively, electrons, photons, and gravitons, while dashed lines denote the charged scalar field.}
	\label{fig05}
\end{figure}

We consider elastic scattering between a charged scalar and an electron, $\phi(p_{1})\,e^{-}(p_{2})\to \phi(p_{3})\,e^{-}(p_{4})$, mediated by photon exchange and by the agravity (see Fig.~\ref{fig05}). The unpolarized squared matrix element is obtained by summing over the final-state electron spin and averaging over the initial-state spin,
\begin{equation}
\overline{|\mathcal{M}|^{2}}
=
\frac{1}{2}\sum_{s,s'}
\left|\mathcal{M}_{\,s\to s'}\right|^{2}.
\label{eq:polavg_sese}
\end{equation}

In terms of the Mandelstam invariants $s,t,u$ (with $s+t+u=0$ in the massless limit), the
unpolarized squared matrix element can be written as
\begin{equation}
\overline{|\mathcal{M}|^{2}}
=
-\frac{s\,u}{16\,t^{4}}
\Big[
f_{2}^{2}\,(s-u)+8\,e^{2}\,t
\Big]^{2}.
\label{eq:M2_sese}
\end{equation}
Several features of Eq.~\eqref{eq:M2_sese} deserve emphasis. First, the explicit $t^{-4}$ dependence shows that the dominant angular enhancement is controlled by the $t$-channel momentum transfer, $t=(p_{2}-p_{4})^{2}$: in the forward limit $t\to0$ the amplitude exhibits a strong IR (collinear) amplification characteristic of massless exchange in a higher-derivative gravitational sector. Second, the squared combination in brackets makes manifest the decomposition into the pure spin-2 agravity contribution (associated with the $f_{2}$ mode), the purely electromagnetic contribution, and their interference at the amplitude level. Finally, the overall prefactor $-s u$ ensures positivity in the physical scattering region, where $s>0$ and $u<0$, so that $-s u>0$ and therefore $\overline{|\mathcal{M}|^{2}}\ge 0$.

In the CoM frame, the result can be expressed as a function of the scattering angle $\theta$,
\begin{equation}
\overline{|\mathcal{M}|^{2}}_{\rm CoM}(\theta)
=
\frac{1+\cos\theta}{8\,(\cos\theta-1)^{4}}
\Big[
f_{2}^{2}\,(3+\cos\theta)
+8\,(\cos\theta-1)\,e^{2}
\Big]^{2}.
\label{eq:M2_sese_CoM}
\end{equation}
Equation~\eqref{eq:M2_sese_CoM} makes the forward enhancement manifest: as $\theta\to 0$ one has $\cos\theta-1\simeq -\theta^{2}/2$, so the prefactor $(\cos\theta-1)^{-4}$ yields a steep small-angle growth. This behavior is the angular avatar of the $t^{-4}$ dependence in Eq.~\eqref{eq:M2_sese} and reflects the sensitivity of the $t$-channel exchange to arbitrarily small momentum transfer in the massless limit. In practice, physically meaningful integrated observables require the standard IR definition through an angular cut (equivalently, a cut on $|t|$) determined by experimental resolution or by the relevant IR physics.

\begin{figure}[h!]
	\includegraphics[angle=0 ,width=7cm]{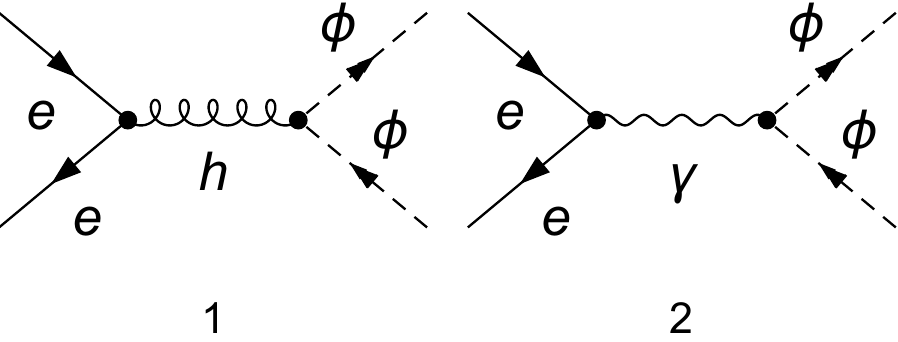}
	\caption{Tree-level diagrams for the annihilation channel $e^-e^+ \to \phi^\dagger\phi$, including photon exchange and graviton exchange. 
    }
	\label{fig06}
\end{figure}

The unpolarized squared matrix element for the annihilation channel $e^{+}e^{-}\to\phi^{\dagger}\phi$, Fig. \ref{fig06}, reads
\begin{equation}
\overline{|\mathcal{M}|^{2}}_{\,e^{+}e^{-}\to\phi^{\dagger}\phi}
=
\frac{t\,u}{16\,s^{4}}
\Big[
f_{2}^{2}\,(t-u)-8\,e^{2}\,s
\Big]^{2},
\label{eq:M2_epss}
\end{equation}
where the bar indicates the standard sum over final-state quantum numbers together with the average over the initial-state spins. In the CoM frame, the kinematics may be parametrized by the scattering angle $\theta$ between the incoming electron and the outgoing scalar, and Eq.~\eqref{eq:M2_epss} reduces to the compact angular form
\begin{equation}
\overline{|\mathcal{M}|^{2}}_{\rm CoM}(\theta)
=
\frac{\sin^{2}\!\theta}{64}\,
\Big(f_{2}^{2}\cos\theta-8e^{2}\Big)^{2}.
\label{eq:M2_epss_CoM}
\end{equation}
Equation~\eqref{eq:M2_epss_CoM} makes explicit the characteristic $\sin^{2}\!\theta$ suppression in the forward and backward directions, $\theta\to 0,\pi$, together with a nontrivial interference pattern between the electromagnetic contribution and the spin-2 agravity interaction encoded in the combination $f_{2}^{2}\cos\theta-8e^{2}$.

\subsection{Compton scattering of a charged scalar: $\gamma\phi \to \gamma\phi$}\label{subsec:Compton_gphi}

\begin{figure}[h!]
	\includegraphics[angle=0 ,width=14cm]{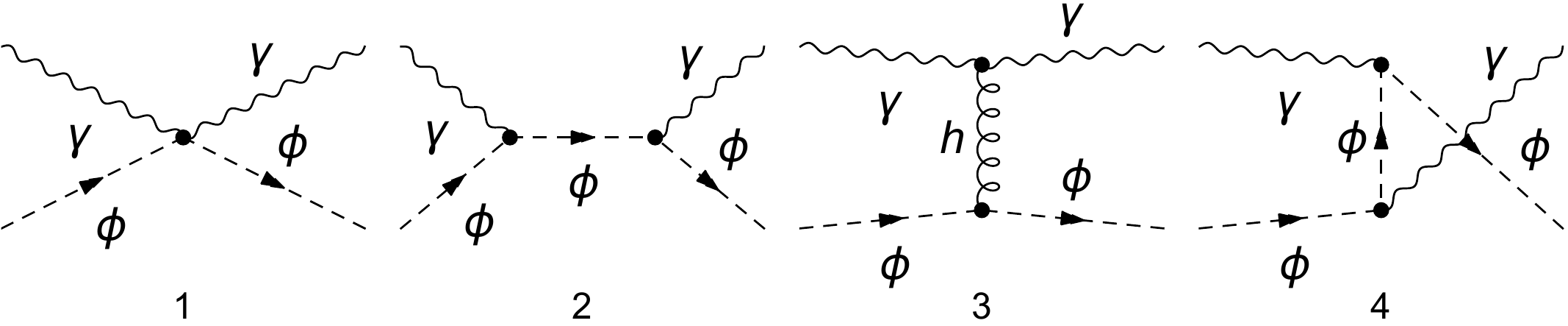}
    \caption{Tree-level Feynman diagrams for $\gamma\,\phi \to \gamma\,\phi$ (scalar Compton scattering), showing the photon-exchange contribution and the graviton-exchange contribution.}
	\label{psps-scat}
\end{figure}

We consider the elastic Compton-like process
\(\gamma(p_{1})\,\phi(p_{2}) \to \gamma(p_{3})\,\phi(p_{4})\),
mediated by photon exchange and by the spin-2 sector of agravity, see Fig. \ref{psps-scat}.  Throughout this subsection we assume an unpolarized photon beam; the sum over final-state photon helicities and the average over initial-state helicities have been implemented, yielding the usual overall factor $1/2$ for the initial photon polarization average.

In terms of the Mandelstam invariants \(s,t,u\) (with \(s+t+u=0\) in the massless limit), the unpolarized squared matrix element takes the compact form
\begin{equation}
\overline{|\mathcal{M}|^{2}}
=
\frac{\big(f_{2}^{2}\,s\,u+4e^{2}\,t^{2}\big)^{2}}{4\,t^{4}}\,.
\label{eq:M2_gphi_to_gphi}
\end{equation}
Equation~\eqref{eq:M2_gphi_to_gphi} makes manifest the decomposition into the pure spin-2 agravity contribution, the purely electromagnetic contribution, and their interference at the amplitude level.  Moreover, the overall $t^{-4}$ dependence shows that the dominant enhancement is governed by the $t$-channel momentum transfer, $t=(p_{1}-p_{3})^{2}$: in the forward-scattering regime $t\to0$ the squared amplitude exhibits a strong collinear amplification, characteristic of massless exchange in a higher-derivative gravitational sector.

In the CoM frame, the result depends only on the scattering angle $\theta$ between the incoming photon and the outgoing photon. Using the standard massless relations, Eq.\eqref{eq:stu_com_massless}, one finds
\begin{equation}
\overline{|\mathcal{M}|^{2}}_{\rm CoM}(\theta)
=
\frac{\Big[f_{2}^{2}\,(1+\cos\theta)-8e^{2}\,\sin^{4}\!\big(\tfrac{\theta}{2}\big)\Big]^{2}}
{(\cos\theta-1)^{4}}\,.
\label{eq:M2_gphi_to_gphi_CoM}
\end{equation}
The factor $(\cos\theta-1)^{-4}$ makes explicit the steep forward enhancement: as
\(\theta\to 0\) one has \(\cos\theta-1\simeq -\theta^{2}/2\), and therefore
\(\overline{|\mathcal{M}|^{2}}_{\rm CoM}\propto \theta^{-8}\) up to the angular dependence contained
in the bracket. 

\subsection{The scalar-scalar scattering}

\begin{figure}[h!]
	\includegraphics[angle=0 ,width=16cm]{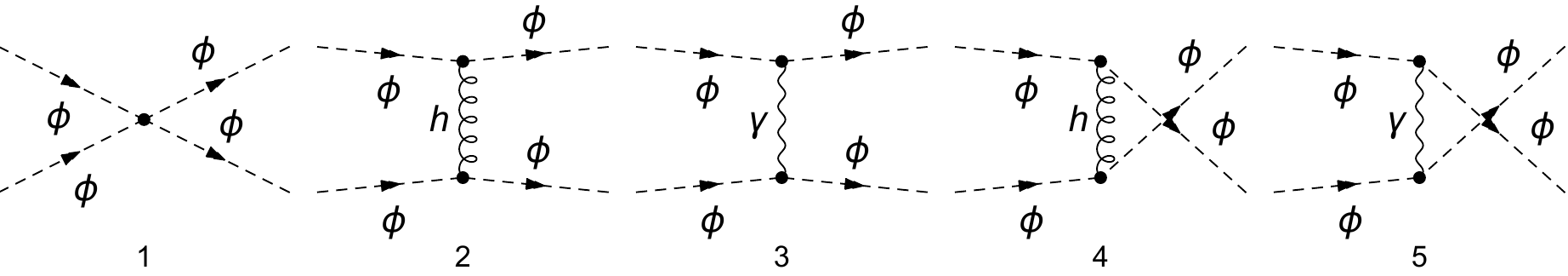}
    \caption{Tree-level diagrams contributing to $\phi\,\phi \to \phi\,\phi$ (elastic charged scalar scattering), including the quartic scalar interaction, photon exchange, and graviton exchange.}
	\label{fig07}
\end{figure}

We now consider elastic scalar scattering, $\phi\,\phi \to \phi\,\phi$, mediated by the quartic self-interaction, photon exchange, and the agravity (see Fig.~\ref{fig07}). After summing over final-state quantum numbers and adopting the standard unpolarized normalization (for scalars this amounts simply to the squared amplitude itself), we obtain the compact expression
\begin{eqnarray}
\overline{|\mathcal{M}|^{2}}
&=&
\frac{1}{36\,t^{4}u^{4}}
\Big[
\big(6\lambda+f_0^2(1+18\,\xi)^{2}\big)t^{2}u^{2}
-f_{2}^{2}\Big(3s^{2}(t+u)^{2}+t u\,(3t^{2}+7tu+3u^{2})\Big)
\nonumber\\
&&\qquad +12\,e^{2}\,t u\,(t^{2}+tu+u^{2})
\Big]^{2}.
\label{eq:M2_ssss}
\end{eqnarray}
Here $\lambda$ is the scalar quartic coupling, $e$ is the Abelian gauge coupling, $\xi$ is the nonminimal curvature coupling appearing in $\xi\,\phi^{\dagger}\phi\,R$, and $f_{0},f_{2}$ are the dimensionless gravitational couplings controlling the spin-0 ($R^{2}$) and spin-2 (Weyl-squared) sectors of agravity, respectively. The structure of Eq.~\eqref{eq:M2_ssss} makes explicit that the result organizes itself as the square of a single kinematic polynomial, exhibiting the interference among the contact interaction ($\lambda$ and $f_{0},\xi$), the spin-2 exchange ($f_{2}$), and the gauge contribution ($e$). The overall factor $t^{-4}u^{-4}$ signals the strong enhancement in the regions where the momentum transfer in the $t$- or $u$-channel becomes small; as usual for massless exchanges, fully integrated observables require an angular (or $|t|,|u|$) cut set by the IR definition of the process.

In the CoM frame, the squared amplitude becomes a function of the scattering angle $\theta$,
\begin{eqnarray}
\overline{|\mathcal{M}|^{2}}_{\rm CoM}(\theta)
&=&
\frac{\csc^{8}\!\theta}{2304}
\Bigg[
3\Big(6\lambda+f_{0}^2(1+18\,\xi)^{2}-81f_{2}^{2}\Big)
-4\Big(6\lambda+(f_{0}+18\,\xi\,f_{0})^{2}+35f_{2}^{2}\Big)\cos 2\theta
\nonumber\\
&&\hspace{1.35cm}
+\Big(6\lambda+f_{0}^2(1+18\,\xi)^{2}-f_{2}^{2}\Big)\cos 4\theta
+48\,e^{2}\,(7+\cos 2\theta)\,\sin^{2}\!\theta
\Bigg]^{2}.
\label{eq:M2_ssss_CoM}
\end{eqnarray}
Equation~\eqref{eq:M2_ssss_CoM} makes the forward/backward enhancement manifest through the factor $\csc^{8}\!\theta$, which reflects the combined $t$- and $u$-channel kinematics in massless
$2\to2$ scattering. The remaining trigonometric polynomial in brackets encodes the detailed pattern of interference among the scalar contact sector, the electromagnetic exchange, and the spin-2 agravity contribution.

It is worth clarifying why the scalar (spin-0) component of the quadratic-gravity propagator does not contribute to several of the massless QED amplitudes considered above, while it does contribute to the scalar--scalar channel. In the weak-field expansion, the graviton couples linearly to the matter energy--momentum tensor through an interaction of the schematic form $h_{\mu\nu}T^{\mu\nu}$. Accordingly, the exchange of the spin-0 mode is governed by the trace $T^\mu{}_\mu$ (and, more generally, by the longitudinal/trace structures that reduce to it after the projector decomposition). For on-shell massless external fermions and photons, $T^\mu{}_\mu$ vanishes (up to total derivatives and/or terms proportional to the classical equations of motion), so that the spin-0 exchange either identically drops out or reduces to contactlike contributions that vanish between physical external states. In contrast, for a scalar field with a nonminimal coupling $\xi\,R\,\phi^\ast\phi$, the improved energy--momentum tensor contains an extra $\xi$-dependent piece whose trace does not vanish on shell in general, thereby activating the spin-0 channel and yielding explicit $f_0$-dependent contributions in the corresponding amplitudes.

\section{Final remarks}
\label{summary}

In this work we have presented a systematic set of tree-level $2\to2$ scattering amplitudes in agravity coupled to an Abelian gauge theory with charged scalars and fermions. Our emphasis has been on compact analytic expressions for the unpolarized squared matrix elements, written in terms of the Mandelstam invariants and, when convenient, in the CoM frame. In contrast with treatments restricted to purely graviton-mediated channels, we have retained throughout the mixed photon--graviton interference contributions, thereby making explicit how the higher-derivative gravitational sector reshapes familiar QED scattering once standard gauge interactions are simultaneously present. We have also verified, by keeping an arbitrary gravitational gauge-fixing parameter $\zeta_g$, that the final results are $\zeta_g$-independent, as required by diffeomorphism gauge invariance at tree level.

A recurrent feature of the amplitudes is a pronounced IR/collinear enhancement in the forward and backward directions. In agravity this behavior is naturally strengthened by the four-derivative graviton propagator, which leads to higher powers of $t^{-1}$ and/or $u^{-1}$ in the corresponding channels. The resulting small-angle structure should therefore be viewed as the long-distance imprint of massless exchange in a higher-derivative gravitational theory, rather than as an UV pathology. Consequently, predictions for integrated observables must be formulated with the usual IR definition of the process---through angular cuts, finite-resolution prescriptions, or appropriate infrared regulators---closely paralleling the standard treatment in gauge theories, albeit with a parametrically stronger sensitivity to the small-angle region.

It is also useful to separate angular dependence from energy scaling in the ultra-Planckian regime. When expressed in the CoM frame, the squared amplitudes become functions of the scattering angle alone, $\overline{|\mathcal{M}|^{2}}_{\rm CoM}=\overline{|\mathcal{M}|^{2}}_{\rm CoM}(\theta)$, with no residual dependence on the total energy $s$ at fixed dimensionless couplings. This follows directly from dimensional analysis in a classically scale-free theory: at tree level the matrix elements depend only on ratios of Mandelstam invariants, so that upon using the massless CoM relations \eqref{eq:stu_com_massless} all explicit powers of $s$ cancel inside $\overline{|\mathcal{M}|^{2}}$. The energy dependence reappears, as it must, when forming the differential cross section,
\begin{equation}
\frac{d\sigma}{d\Omega}
=
\frac{1}{64\pi^{2}s}\,\overline{|\mathcal{M}|^{2}}_{\rm CoM}(\theta),
\label{eq:dsigdOmega_scaling}
\end{equation}
so that the angular profile is entirely dictated by $\overline{|\mathcal{M}|^{2}}_{\rm CoM}(\theta)$, while the overall normalization decreases with energy through the universal phase-space factor.

At ultra-Planckian energies the amplitudes display the scaling expected of a renormalizable, scale-free gravitational sector, together with nontrivial patterns of cancellation and interference among gravitational and electromagnetic contributions. In particular, the interference terms provide a sensitive diagnostic of how the spin-2 sector (controlled by $f_{2}$) modifies QED kinematics across channels involving photons, fermions, and scalars. In this sense, the results collected here extend recent amplitude-based studies of quadratic gravity to a broader set of matter processes and offer a concrete arena in which to examine, channel by channel, the interplay between higher-derivative gravitational dynamics and conventional gauge interactions.

Several natural extensions suggest themselves. First, it would be interesting to incorporate the running of the dimensionless couplings $(f_{2},f_{0},e,\lambda,\xi)$ and to study how renormalization-group evolution reshapes the relative importance of pure-gravity and interference terms across kinematic regimes. Second, while tree-level amplitudes already capture the dominant IR structure, inclusive observables at high energy ultimately require the standard treatment of soft/collinear radiation and loop corrections; implementing this program in agravity--QED would clarify how higher-derivative gravity interfaces with the familiar IR factorization of gauge theories. Finally, the analytic expressions derived here may serve as building blocks for phenomenological estimates and for further comparisons with the non-Abelian sector, alongside our earlier ultra-Planckian analysis in agravity-coupled QCD.

\bigskip

\acknowledgments
I.F.C. is partially supported by Coordena\c{c}\~ao de Aperfei\c{c}oamento de Pessoal de N\'ivel Superior (CAPES). The work of A.C.L. was partially supported by Conselho Nacional de Desenvolvimento Cient\'{i}fico e Tecnol\'{o}gico (CNPq), Grants No.~404310/2023-0 and No.~301256/2025-0.

\vspace{1cm}

\end{document}